# Cooperation and competition in the dynamics of tissue architecture during homeostasis and tumorigenesis

Attila Csikász-Nagy[1,2], Luis M. Escudero[3], Martial Guillaud[4], Sean Sedwards[5], Buzz Baum[6], Matteo Cavaliere[7]

[1] Department of Computational Biology, Research and Innovation Center, Fondazione Edmund Mach, San Michele all'Adige 38010, Italy

[2] Randall Division of Cell and Molecular Biophysics and Institute for Mathematical and Molecular Biomedicine, King's College London, London SE1 1UL, United Kingdom

[3] Instituto de Biomedicina de Sevilla. Hospital Universitario Virgen del Rocío/CSIC/Universidad de Sevilla, Seville, Spain.

[4] Department of Integrative Oncology , British Columbia Cancer Research Center, Vancouver, British Columbia, V5Z 1L3 Canada

[5] INRIA Rennes – Bretagne Atlantique, Campus universitaire de Beaulieu, 35042 Rennes Cedex, France

[6] MRC Laboratory for Molecular Cell Biology, University College London, Gower Street, London WC1E 6BT, United Kingdom

[7] Logic of Genomic Systems Laboratory, Spanish National Biotechnology Centre (CNB-CSIC), Madrid 28049, Spain

**Corresponding authors:**

Attila Csikász-Nagy, attila.csikasz-nagy@fmach.it, tel: +39 0461615645, fax: +39 0461615416

Matteo Cavaliere, mcavaliere@cnb.csic.es

**Abstract**

The construction of a network of cell-to-cell contacts makes it possible to characterize the patterns and spatial organisation of tissues. Such networks are highly dynamic, depending on the changes of the tissue architecture caused by cell division, death and migration. Local competitive and cooperative cell-to-cell interactions influence the choices cells make. We review the literature on quantitative data of epithelial tissue topology and present a dynamical network model that can be used to explore the evolutionary dynamics of a two dimensional tissue architecture with arbitrary cell-to-cell interactions. In particular, we show that various forms of experimentally observed types of interactions can be modelled using game theory. We discuss a model of cooperative and non-cooperative cell-to-cell communication that can capture the interplay between cellular competition and tissue dynamics. We conclude with an outlook on the possible uses of this approach in modelling tumorigenesis and tissue homeostasis.

**Epithelial organisation and cellular interaction networks**

In a monolayer epithelium, apical cell surfaces appear polygonal in shape. In a normal tissue, these polygons fit perfectly together to form a tiled array that is free from gaps. Therefore, for each polygon the number of sides corresponds to the number of cell neighbours in the tissue. By using this information to construct an "epithelial network" of cell-to-cell contacts (or cellular interaction network) it is then possible to capture information about the patterns in the spatial organization of epithelial cells (Figure 1). In such a network, the centroid of each cell in the epithelium is treated as a node and two nodes are linked if the two cells are neighbours (*i.e.*, they are in physical contact). This approach can help us to understand the local interactions - also termed "cellular sociology" [1] - in tissues.

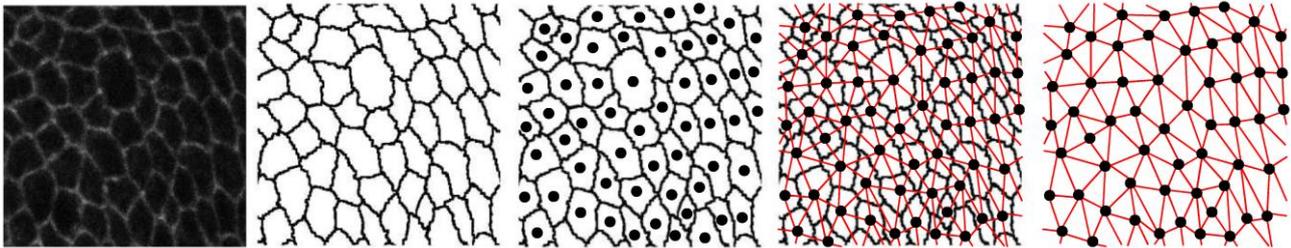

**Figure 1. From tissue topology to the cellular interaction network.** The panels show the different processing step from left to right. The original image (*Drosophila* wing disc in this case) is segmented to obtain the outline of all cells. This allows the localization of the centroid (nodes) and the identification of the neighbours (links) of every cell, the basis of the cellular interaction network. The contacts network is called Delaunay graph, while the cell outline graph is called Voronoi diagram.

Conceptualising the arrangement of cells as a network opens up new possibilities to investigate tissue organization and patterns using principles from graph and network theory [2]. For instance, when considering the cellular interaction network of the epithelial tissue of the developing *Drosophila* wing [3] one sees that while most cells have six direct neighbours, some have four or nine neighbours [4]. Remarkably, other multicellular tissues show very similar distribution of sidedness (number of neighbours per cell or 'degree' of nodes in the contacts networks) [4, 5].

Several theoretical hypotheses have been proposed to explain how this observed distribution could emerge, using elasticity theory [6], mechanics [7, 8], or stochastic models of cell replication [4, 9]. However, while the degree distributions seem to follow a general principle [4] that can be captured by the proposed mathematical models, more specific topological measures of cellular contact networks, e.g. the degree or the cluster coefficient of nodes (a proxy of the connectedness density), can be used to distinguish both tissues and species [2, 10]. Specifically, the quantitative signature of a tissue can be obtained by considering an appropriate set of statistics concerning the cellular contacts networks [10]. For instance, the analysis of muscle biopsies allows the evaluation of differences between normal and pathological muscle tissues. This method not only automates the work of the pathologist by capturing the geometrical information from the biopsy, but also captures the spatial organization of muscle fibres by constructing a "muscle network" of fibre-to-fibre contacts. Interestingly, using this approach some "network features" (among a list of 82 possible features) were found as key parameters to quantify the degree of pathology of muscular dystrophy patients [10]. One can therefore conclude that the topological organization of the epithelium differs between tissues and between species and can provide valuable information. On the other hand, the highly reproducible network structure of the same epithelia taken from different individuals hints at the presence of genetic controls that guide local and long-range tissue organization, such as force-dependent cell division and cell competition [11], suggesting multiple levels of control which shape the overall cellular interaction network of living tissue across multiple scales.

**The forces that shape epithelial topology**

The general phenomenon termed as cell competition can arise when two populations of cells, differing in their growth, coexist in a developing tissue. In a homotypic population, cells will proliferate and the tissue will develop normally. However, when two populations coexist in a tissue and must compete for space, the fitter (faster growing) of the population will proliferate at the expense of the less fit population. The "winner" cells can eventually take over the developmental compartment by inducing apoptosis in "loser" cells [12, 13]. In addition, identical cells have been shown to compete for space in crowded epithelia in different model organisms so that 'loser' cells are lost by delamination [14, 15]. Thus, mechanical interactions and molecular signalling (communication) between surrounding cells can drive localized cell death in epithelia. In healthy tissues the removal of loser cells and winner proliferation appear to be carefully balanced, such that the competition acts to return a tissue to the homeostatic condition. Conversely, it has been proposed that cancer cells, which lack the controls necessary to respond to tissue overcrowding, out-compete their neighbouring wild-type cells through a process known as field cancerization, leading to the growth of the early-stage hyperplastic tumour at the expense of the host tissue [16, 17]. Specifically, epithelial pre-neoplastic lesions progress from normal homeostatic differentiated epithelium to undifferentiated carcinoma in situ through a complex accumulation of genetic mutations associated to loss of tissue architecture and modification of cellular interactions. It is therefore important to determine how tissue topology changes in different epithelial tissues such as mucosa, skin, lung or cervix.

Technically, the tissue topology can be obtained by analysing patient histological samples with algorithms from graph theory, such as Voronoi diagrams, Delaunay triangulations, Minimum Spanning Tree, or other neighbourhood-based distance algorithms, [18]; for instance, these approaches are used to assess the degree of epithelial differentiation, loss of order and homeostatic equilibrium in epithelial pre-neoplastic lesions [19] as well as loss of cell differentiation in carcinomas [20]. Statistical parameters of the cellular interaction networks combined with precise measurement of morphological changes in cell nuclei have shown promising potential as surrogate markers for malignancy grade and cancer progression [21, 22]. Figure 2 presents different examples of p16 immuno-stained cervical pre-neoplastic lesions and the corresponding obtained tissue topology network. Recent advances in imaging, microscopy, optical technologies, and digital pathology, make possible high-throughput analysis and simultaneous measurements of multiple proteins, and other molecules (miRNA, etc.) in histological specimens and tissue microarrays. This allows the segmentation and identification of subpopulations (clones) of genetically similar cells within tissue samples through measurement of loci-specific fluorescence *in-situ* Hybridization (FISH) spot signals for each nucleus [23]. These new methodologies can then facilitate the construction of the global topology of an epithelial tissue and the quantitative analysis of the spatial distribution of clonal subpopulations.

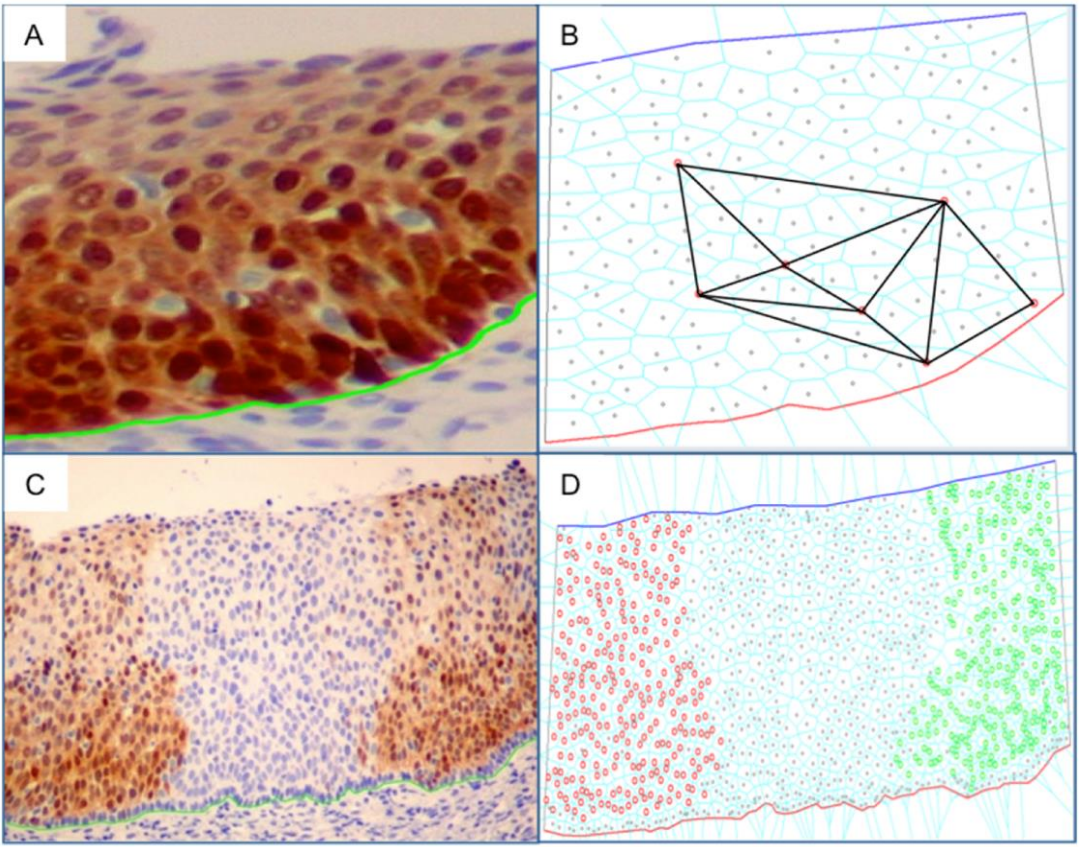

**Figure 2**. **Analysis of p16 expression in cervical pre-neoplastic lesions.** p16 is a protein over-expressed in HPV infected cells. Chronic infection with HPV ultimately causes cervical cancer, but fortunately only a small number of HPV infected epithelium will progress to carcinoma. (**A**) Isolated HPV negative cells in a strong HPV positive (brown stain) field with corresponding topological graph (**B**). (**C**) Well demarcated HPV positive fields and Voronoi diagram (**D**) used as cell outline graph.

**Cellular interactions and game theory**

Cell competition, cooperation and more general cellular interactions present in living tissues can be studied using game theory [24-27]. Game theoretical approaches are widely used to analyse interactions between individuals using different strategies. In evolutionary game theory one assumes that strategies ("phenotypes") are associated with genotypic variants, enabling one to analyse the emergence and spreading of specific successful strategies in a population [28]. Individuals using successful strategies will have higher fitness, reproduce faster and spread in the population. One of the most analysed evolutionary games, known as the prisoner's dilemma, is used to capture the notion of cooperation and defection (cheating) (Figure 3). In this game a cooperator pays a cost to distribute benefits to its interaction partners. Defectors (cheaters) do not pay any cost and do not yield benefits to their partners. If benefit and costs result in corresponding changes to fitness, then a well-mixed population of cooperators can be invaded by cheaters. The spreading of cheaters leads to a decrease in the average population fitness and can lead to population collapse if its sustainability depends on the presence of cooperators (in what is known as the "tragedy of the commons") [29].

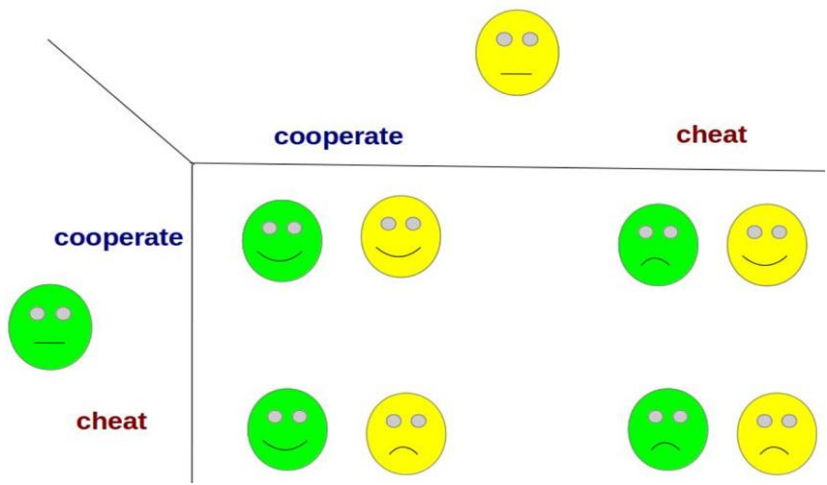

**Figure 3. The Dilemma of Cooperation.** The prisoner's dilemma is a game to analyze the problem of cooperation between individuals that interact. Individuals can choose between two strategies: to cooperate or to cheat. The best payoff is obtained when both individuals cooperate, however the temptation of cheating (obtaining a better payoff than the opponent) can lead both individuals to cheat, leading to the worst outcome. In evolutionary game theory payoffs are interpreted as fitness: Individuals using successful strategies replicate faster. The game can be generalized to any population of interacting individuals.

Evolutionary game theory has been used to model the evolution of specific strategies in various complex networks [30-34] as well as in population of spatially structured cells [24, 26, 30, 33, 35-38]. This has been generally done by arranging cells on a static grid (e.g., cellular automaton) and following how their interactions influence their movement and proliferation [39]. This approach can be extended to study tissue topologies, but the fixed grid puts an unrealistic limit on the dynamical changes of the cellular interaction network. This is evident when the abstract notion of cooperation and cheating among cells is mechanistically explained by different types of cell-to-cell communication. Many distinct mechanisms are known to exist. For instance, direct membrane-bound signalling-receptor interactions can send information between cells to drive competition for cell fate [40] or to repel dissimilar cells [41]. This can even occur over long distances [40, 42] so that signalling molecules on the surface of one cell are detected by receptors on the surface of non-nearest neighbours over distances of many microns to alter their behaviour. These receptors induce specific responses in the receiver cell, which can depend on the concentration and the duration of a signal, and may lead to the expression of another signalling molecule. In this light, cooperative cells are those that actively engage in communication (with the "cost" of producing and listening to the communication signals) and that help to shape the local tissue environment, [40, 42]. This is a similar to the presence of mechanical forces that can alter the division rate or death rate of cells, [15, 43, 44], locally or at a distance.

In this context, a defecting cell is one that has lost some of the controls from its environment that enable it to behave differently, e.g., by ignoring some of the signalling inputs. In a cellular prisoner's dilemma, cheating cells that do not signal and/or ignore signals have less costs than cooperative cells and can replicate faster, ultimately leading to a tissue where the global signalling system is compromised. In all cases, signalling among the cells is strictly linked to the tissue architecture, hence to the described cellular interaction networks. The interactions alter the cell's growth and death rates depending on the cells' neighbours. This influences tissue topology which, ultimately, feeds back to influence the microenvironmental cell-to-cell communication. Because of this, cellular interaction networks and cellular communications are interrelated and highly dynamic.

**Modelling cell-to-cell interactions using game theory and dynamical networks**

To investigate the described complex interplay we consider cell replication and cell death in the framework of dynamical networks and game theory [30, 45]. A dynamical network can model the cellular interaction network of an epithelial tissue changing in time, depending on the interactions (the "game") between adjacent cells and the resulting localized cell replication and death. Such a framework [30, 45] considers how tissue topology is altered by stochastic cell duplication and death events that occur with probabilities determined by averaging the payoff of a game played between neighbouring cells. Cells then compete for resources and those with higher fitness can replicate faster. Healthy tissues can be modelled as a uniform population of cooperative cells. In fact, the polygon distribution of uniformly cooperating cells in the model match those of healthy biological tissues. When the network contains a sufficient number of defectors the topology is different, modelling some form of pathology (Figure 4). This simple model reveals the profound influence of cell competition on tissue topology [30, 45].

**Cell competition, epithelial organization and development of cancer**

Cancers evolve over time through the mutation and selection of individual cells in the population and can therefore be viewed as adaptive Darwinian systems [46, 47]. Despite the differences between cancer types, one can generalize see solid tumours arising as result of the "evolution of… defection and the breakdown of cooperation" [48] . Following the prisoner's dilemma metaphor, cancerous cells are viewed as cheaters. In a healthy tissue, cells replicate in an organized manner as requested by the developmental program. Cooperative cells are those that produce and respond to the crosstalk of signalling molecules coordinating the developmental program and indicating whether a cell can divide or should go to apoptosis. Cheating (mutated or cancerous) cells do not participate to such signalling system, obtain a fitness benefit and can divide in a selfish, uncontrolled, way.  As for other cells, however, the reproductive ability of cancerous cells is shaped by their interactions with neighbouring cells and their microenvironment [49-51] and tissue architectures likely evolved to optimize multicellular functions and restrain the spreading of cheating cells. This resilience of a healthy tissue may then depend on the spatial organisation and specific tissues structures (hence on the cellular interaction network). This has been shown using mathematical models where the somatic evolution of cancer is considered in a linear array of cells [52]. This idea can be generalized in the proposed dynamical networks framework. One can consider an arbitrary two dimensional tissue in which cooperative and defective cells can divide or be removed by apoptosis, and where the fitness of each cell is shaped by its local interactions, [30, 45]. The dynamical tissue structure and the differential fitness of cooperative and defective cells – modelling cell competition - leads to a complex evolutionary and topological phenomena: defecting cells are fitter than adjacent cooperative cells but homotypic populations of defecting cells are disadvantaged, as there is little benefit from interactions with only defecting cells. Individual defecting cells embedded in a population of cooperative cells however have an advantage and can proliferate faster than the surrounding tissue. As the population of defecting cells grows, however, it will reach a limiting size as defectors become surrounded by other defectors, leading to a fitness disadvantage. This leads to isolated populations of defectors surrounded by cooperative cells that could be described as benign tumour-like (Figure 5A). Further increases in the defector population can happen only with elevated cell death, as can occur if loser cells in contact with defectors have a higher chance of removal (Figure 5B), as seen in some cancer models [13]. The cell interaction network of the defecting population is very different from that of cooperators tissue in which is embedded and, strikingly, similar changes in tissue cytoarchitecture have been observed in various tumours [53]. Such examples suggest that changes in tissue architecture influence the process of tumorigenesis.

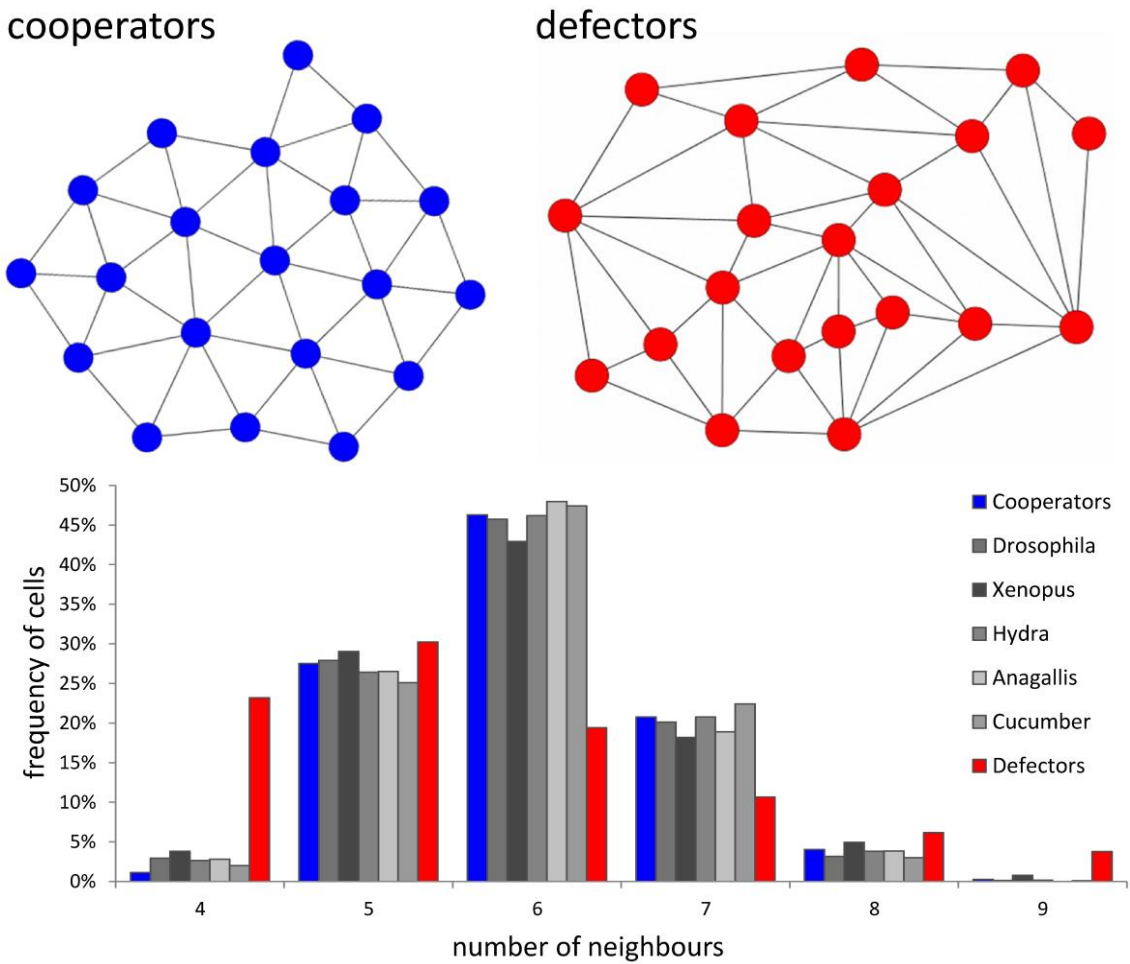

**Figure 4. Topologies of cooperator and defector networks and their relation to epithelial tissues.** Examples of small parts of a larger simulated cellular interaction network of cooperators and defectors and the distribution of sidedness observed in these simulated tissues and in epithelial tissues from various organisms [4].

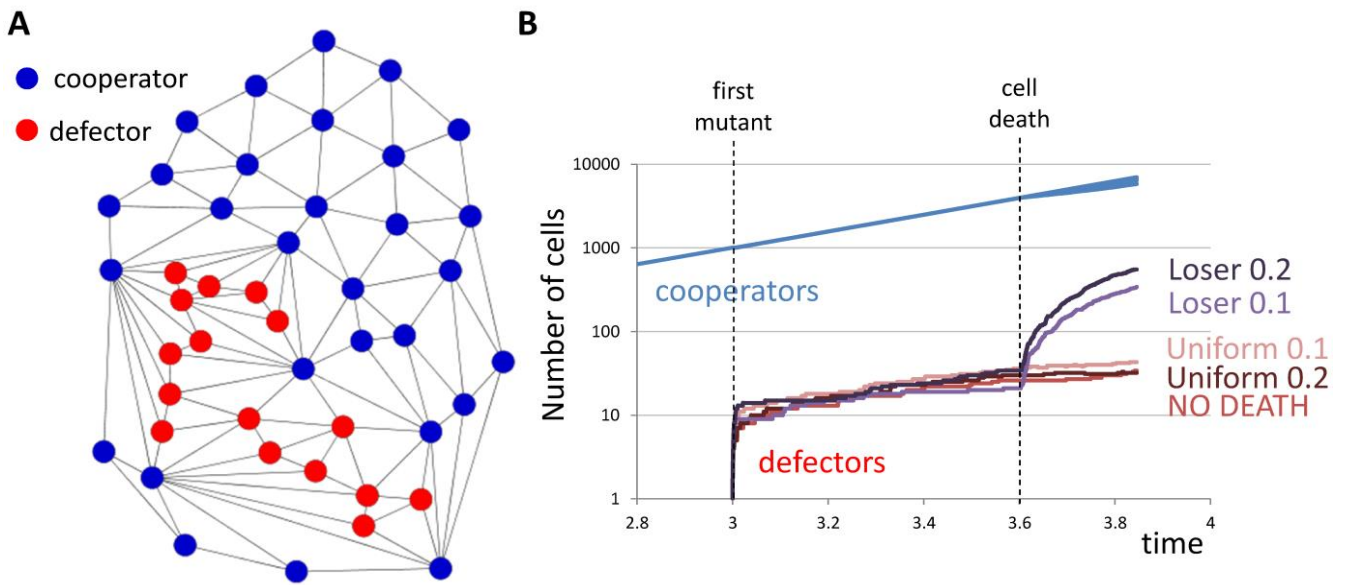

**Figure 5. Spreading of defectors in a cooperating tissue.** A single defecting cell appears in a tissue of 1000 cooperators at time 3 (measured in arbitrary time units here). Without cell death the defector population cannot spread above 40-50 even in a long time scale. Example of such a network on panel A. If cells die (initiated at time 3.6) with equal probability (uniform death at rate 0.1 or 0.2) no further spreading of defecting cells is observed. Contrary, when losers (low fitness cells) have a higher chance to die then defectors can spread further (both at rate 0.1 or 0.2 - corresponding to 10 and 20% chance of cell death for one division step). Note the log scale on the ordinate. The used software tool is available at: *www.dynamicalnetworks.org/planar-tissue*

**Perspectives**

The use of game theory and dynamical networks constitute a general framework to analyse the evolutionary dynamics of a cellular interaction network. We suggest that this framework could become a useful new computational approach to study the process of tumorigenesis as a disease of tissue pattern formation [53], and the role of cell-to-cell interactions and competition in the microenvironmental niche of mutated cells [54]. Indeed, in the case of muscle and neuromuscular diseases, it has already been shown that the cellular interaction network can provide valuable biological insights concerning the aetiology of the disease [10]. Since the majority of cancers originate from epithelia, this approach is particularly relevant as an aid to understanding how cells interact in diseased epithelial tissues, and how hyperplasia and competition could alter the epithelial architecture and change the probability of small precancerous clones expanding and invading adjacent tissues [55]. In addition, local cell-to-cell interactions are important in maintaining tissue homeostasis [14, 15] and controlling stem cell pools [56], and could therefore help to throw light on the processes of aging and regeneration. A further challenge is to move the dynamical network analysis to real 3D structures. This could be done by extending the existing modelling platform and tissue sample analysis method used to capture the spreading of a mutant in a 3D population of cells [18]. In fact, this method can be used to create 2D cuts of the simulated tissues and compare them with patient samples. The available experimental analysis of expression and distribution of proteins *in situ* allows us to identify the individual cellular phenotypes present in a tissue (the "strategies" that cells are using). For instance, Figure 2 presents examples of the analysis available for diagnostic areas exhibiting distinct expression patterns. The analysis of the cellular interaction network of various tissues is already achievable [2, 10, 18]. Related approaches are also used to investigate plant development [57, 58] and some of these might be adapted to cancer research as well. At this stage, however, these approaches deal with static pictures because it is technically challenging to follow the changes of cellular interaction networks over time [14, 15]. Further developments along these lines will be necessary, enabling us to follow the interactions between various cell types and the resulting changes in tissue topology.

In our view, the framework bridging evolutionary game theory and dynamical networks can already help us take a step forward using only the available static data. Fitting topologies of various tissue types, i.e., with and without mutated (cancerous) cells, with opportune dynamical networks models could suggest the types of interactions that might have led to the observed tissue topology (e.g., trying to reverse-engineer the types of phenotypic strategies that cells have been using). Moreover, the dynamical network model presented could be then associated with the available analysis of patient samples [10, 18] to provide specific network measures to discriminate the various evolutionary stages of a tissue in which cancerous mutants are spreading. The topological characterization of these stages could be a first step towards understanding what allows some mutants to spread and others to die, and to determine the evolutionary trajectory of the interaction network leading to the collapse of cellular cooperation. Game theory could also help to consider more general types of cell-to-cell interactions in the microenvironmental niche of mutated cells [54] and evaluate the role of tissue topology and cell-to-cell interactions in the homeostatic maintenance of normal cytoarchitecture and the perturbations in it during cancer formation. This could complement the other tissues modelling approaches present in the literature [59-65], especially those that are focusing directly onto cell communication [66, 67] and epithelial organization [68].

Overall, we believe that an integrated framework combining quantitative analysis of tissues samples, dynamical networks and evolutionary game theory could represent a promising approach to explain recent results regarding molecular cell interactions [54], with the possibility to integrate the intra-intercellular signalling with the changes of tissues architecture, providing new directions to multi scale models of cancer [69-71] and ultimately shedding a novel light on the processes of pre-neoplastic growth.

**Acknowledgements**

LME is funded by the Miguel Servet (Instituto Carlos III) program and by the Spanish Ministry of Science (BFU2011-25734). MC is funded by MEC through project BIO2007-63056.